\documentclass[letterpaper]{article} 
\usepackage{aaai25}  
\usepackage{times}  
\usepackage{helvet}  
\usepackage{courier}  
\usepackage[hyphens]{url}  
\usepackage{graphicx} 
\usepackage{natbib}  
\usepackage{caption} 
\usepackage{algorithm}
\usepackage{algorithmic}

\usepackage{newfloat}
\usepackage{listings}
\DeclareCaptionStyle{ruled}{labelfont=normalfont,labelsep=colon,strut=off} 
\floatstyle{ruled}
\newfloat{listing}{tb}{lst}{}
\floatname{listing}{Listing}
\title{Behavior Importance-Aware Graph Neural Architecture Search for Cross-Domain Recommendation}
\author{
    Chendi Ge\textsuperscript{\rm 1}\thanks{This work was done during the author’s internship at Tencent},
    Xin Wang\textsuperscript{\rm 1,\rm 2}\thanks{Corresponding authors},
    Ziwei Zhang\textsuperscript{\rm 1},
    Yijian Qin\textsuperscript{\rm 1},
    Hong Chen\textsuperscript{\rm 1},
    Haiyang Wu\textsuperscript{\rm 3},
    Yang Zhang\textsuperscript{\rm 3}, 
    Yuekui Yang\textsuperscript{\rm 1,\rm 3},
    Wenwu Zhu\textsuperscript{\rm 1,\rm 2}\footnotemark[2]
}
\affiliations{
    \textsuperscript{\rm 1}Department of Computer Science and Technology, Tsinghua University\\
    \textsuperscript{\rm 2}Beijing National Research Center for
Information Science and Technology, Tsinghua University \\
    \textsuperscript{\rm 3}Machine Learning Platform Department, Tencent TEG \\
    
    \{gcd23, qinyj19, h-chen20\}@mails.tsinghua.edu.cn, \{xin\_wang, wwzhu\}@tsinghua.edu.cn, zw-zhang16@tsinghua.org.cn, \{gavinwu, yizhizhang, yuekuiyang\}@tencent.com
}

\usepackage{bibentry}

\usepackage{amssymb}
\usepackage{amsmath}
\usepackage{booktabs}
\usepackage{multirow}
\usepackage{makecell}
\usepackage{multirow}
\usepackage{enumitem}
\usepackage{algorithm}
\usepackage{float}
\usepackage{algorithmic}
\usepackage{subfig}
\usepackage{makecell}

\newcommand{\model}{BiGNAS }
\newcommand{\modelns}{BiGNAS} 

\begin{document}

\maketitle

\begin{abstract}
Cross-domain recommendation (CDR) mitigates data sparsity and cold-start issues in recommendation systems. While recent CDR approaches using graph neural networks (GNNs) capture complex user-item interactions, they rely on manually designed architectures that are often suboptimal and labor-intensive. Additionally, extracting valuable behavioral information from source domains to improve target domain recommendations remains challenging. To address these challenges, we propose \underline{B}ehavior \underline{i}mportance-aware \underline{G}raph \underline{N}eural \underline{A}rchitecture \underline{S}earch (\textbf{\modelns}), a framework that jointly optimizes GNN architecture and data importance for CDR. \model introduces two key components: a Cross-Domain Customized Supernetwork and a Graph-Based Behavior Importance Perceptron. The supernetwork, as a one-shot, retrain-free module, automatically searches the optimal GNN architecture for each domain without the need for retraining. The perceptron uses auxiliary learning to dynamically assess the importance of source domain behaviors, thereby improving target domain recommendations. Extensive experiments on benchmark CDR datasets and a large-scale industry advertising dataset demonstrate that \model consistently outperforms state-of-the-art baselines. To the best of our knowledge, this is the first work to jointly optimize GNN architecture and behavior data importance for cross-domain recommendation.
\end{abstract}

\begin{links}
\link{Code}{https://github.com/gcd19/BiGNAS}
\end{links}

\section{Introduction}
In recent years, the rapid growth of the Internet has led to significant information overload in online environments. To help users efficiently access content that matches their interests and improve retention and conversion rates, recommendation systems have become crucial to web platforms such as e-commerce and content delivery~\cite{zhang2019deep, zhang2023adaptive, wang2024automated}. However, traditional recommendation systems continue to face challenges like data sparsity~\cite{paul2016, guo2017} and cold-start problems~\cite{zhang2019deep}, primarily due to insufficient data. Cross-domain recommendation (CDR)~\cite{hu2018conet, ouyang2020minet, cui2020herograph, zhu2021cross, cao2022disencdr, ning2023multi} has emerged as a promising solution by aggregating user preferences across multiple domains, alleviating these issues.

Most existing CDR algorithms integrate user-item interaction structures~\cite{hu2018conet, ouyang2020minet} to facilitate information transfer between domains, resembling the "cross-stitch" interaction models~\cite{ishan2016} used in multi-task learning. These methods often employ separate neural networks for each domain, limiting their ability to learn global user preferences effectively across domains. Recently, GNN-based CDR methods~\cite{cui2020herograph, cao2022disencdr, ning2023multi} have gained attention by connecting interaction graphs from source and target domains, enabling more effective information transfer and capturing higher-order interaction patterns between users and items across domains.

However, existing GNN-based cross-domain recommendation algorithms still suffer from two key challenges:
\begin{itemize}
\item \textbf{Model adaptability}: Current approaches rely on fixed architectures designed with domain-specific expert knowledge, which is labor-intensive and limits adaptability across different datasets and tasks.
\item \textbf{Negative transfer}: In sparse source domains, methods like attention mechanisms often overemphasize noisy features, resulting in ineffective transfer and suboptimal recommendations in the target domain.
\end{itemize}

To address the challenges of model adaptability and negative transfer in cross-domain recommendation (CDR), we propose \underline{B}ehavior \underline{i}mportance-aware \underline{G}raph \underline{N}eural \underline{A}rchitecture \underline{S}earch (\modelns). \model is a novel framework that jointly optimizes graph neural network (GNN) architecture and data importance for CDR. It introduces two key components: a Cross-Domain Customized Supernetwork and a Graph-Based Behavior Importance Perceptron. The supernetwork, designed as a one-shot, retrain-free module, automatically searches for the optimal GNN architecture tailored to each domain by capturing domain-specific interaction patterns. Meanwhile, the perceptron uses auxiliary learning to dynamically assess the importance of user behaviors in the source domain, allowing the model to prioritize valuable behaviors and improve target domain recommendations.

To achieve end-to-end training, we employ bi-level optimization with implicit gradients, alternately training the two modules. Extensive experiments on both benchmark CDR datasets and a real-world industry advertising dataset demonstrate that \model consistently outperforms state-of-the-art baselines.

\begin{figure*}[htbp]
    \centering
    \includegraphics[width=\linewidth]{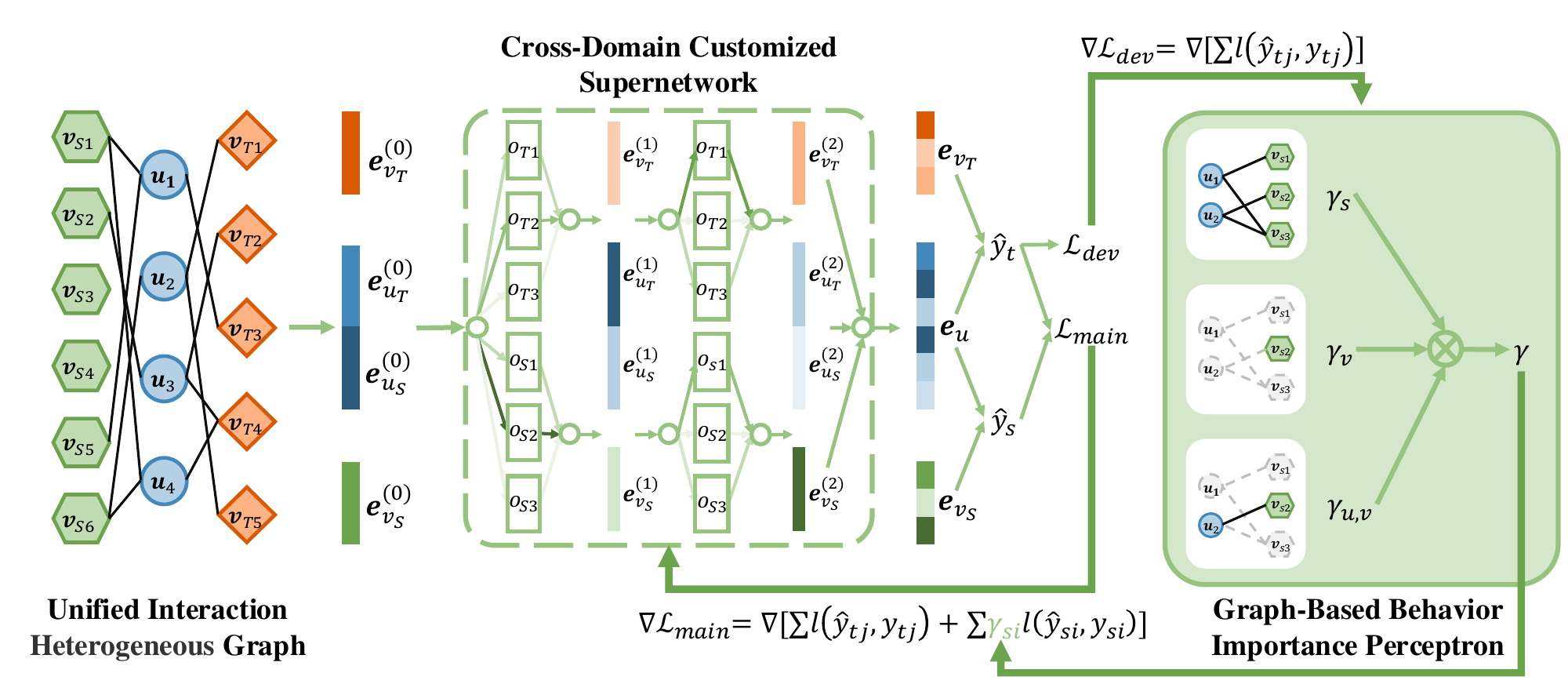}
    \caption{The framework of \model is illustrated using dual-domain CDR. User behavior data from both domains is combined into a unified heterogeneous interaction graph, which serves as the model's input. The model adopts a bi-level structure: the inner Cross-Domain Customized Supernetwork follows a one-shot, retrain-free paradigm to tailor the optimal GNN architecture for each domain, while the outer Graph-Based Behavior Importance Perceptron dynamically evaluates the importance of source domain user interactions through auxiliary learning, guiding model optimization. The two modules are trained alternately to enable end-to-end optimization. The embedding $\mathbf{e}^{(i)}$ is generated by the $i$-th layer of the supernetwork.}
    \label{fig:framework}
\end{figure*}

The contributions of our work are summarized as follows:
\begin{itemize}
    \item We propose a joint optimization framework that autonomously tailors GNN architectures and optimizes data importance for CDR, addressing the unique requirements of each domain. To the best of our knowledge, this is the first work to explore joint optimization of graph neural architecture and data importance in cross-domain recommendation.
    \item We introduce the Graph-Based Behavior Importance Perceptron, which leverages the graph structure of user-item interactions to dynamically adjust the influence of user behaviors during model training, enhancing the effectiveness of cross-domain recommendation.
    \item We conduct extensive experiments on both benchmark CDR datasets and a large-scale industry advertising dataset, demonstrating superior performance compared to existing state-of-the-art methods.
\end{itemize}

\section{Related Works}
\subsection{Cross-domain Recommendation}
Cross-domain recommendation (CDR)~\cite{zhu2021cross} addresses data sparsity and cold-start challenges by aggregating user preferences across different domains. Early CDR methods, like CMF~\cite{singh2008relational}, used matrix factorization (MF) to integrate features across domains but were limited by MF's constraints in modeling user preferences.

Deep learning-based CDR models, such as CoNet~\cite{hu2018conet} and MiNet~\cite{ouyang2020minet}, improved performance by introducing cross-domain interaction via shared projection matrices. However, they remained limited to learning from known user-item pairs and struggled to capture higher-order relationships among users and items.

Graph neural networks (GNNs) have proven effective in overcoming these limitations by modeling complex user-item interactions and higher-order relationships. HeroGRAPH~\cite{cui2020herograph} constructs a heterogeneous graph to represent interactions and leverages global and domain-specific subgraphs for click-through rate prediction. DisenCDR~\cite{cao2022disencdr} further enhances CDR by disentangling domain-shared and domain-specific user representations using a variational bipartite graph encoder and mutual-information regularizers. EDDA~\cite{ning2023multi} is a recent leading CDR method, consisting of an embedding disentangling recommender and a domain alignment strategy.

\subsection{GNNs and Graph Neural Architecture Search}

Graph neural networks (GNNs)~\cite{kipf2017semisupervised, xu2018how, veličković2018graph, li2023curriculum} have proven highly effective in learning node representations via message-passing mechanisms. While their primary applications include node and graph classification, GNNs are also widely used across various domains~\cite{zhang2020deep}. In recommendation systems, where users and items can be modeled as nodes and their interactions as edges, GNNs offer a natural framework for recommendation tasks~\cite{wu2022graph, gao2023survey, wang2023curriculum, zhang2024disentangle1}. GNN-based models, such as NGCF~\cite{wang2019neural} and LightGCN~\cite{he2020lightgcn}, have achieved significant performance gains in these tasks.

The design of neural network architectures remains a challenging and labor-intensive process. Neural Architecture Search (NAS) has emerged as an effective approach to automate this process~\cite{elsken2019neural}. Specifically, graph neural architecture search has gained momentum as a means to automate the design of GNNs~\cite{zhang2021automated, zhang2023dynamic, zhang2023autogt, xie2023adversarially, zhang2024disentangled2, yao2024data, zhang2024unsupervised, xie2024towards, qin2023multitask, cai2024multimodal}. GraphNAS~\cite{gao2020graph}, a seminal method, defines a search space encompassing diverse architectural components and employs a recurrent neural network controller. GNAS~\cite{cai2021rethinking}, a one-shot approach, emphasizes feature filtering and neighborhood aggregation, leveraging DARTS~\cite{liu2018darts} to optimize both weights and architecture. PAS~\cite{wei2021pooling} targets graph classification tasks, offering a search space that integrates aggregation, pooling, readout, and merge functions, along with a coarsening strategy to expedite the search. GASSO~\cite{qin2021graph} simultaneously optimizes graph structure and GNN architecture, whereas GAUSS~\cite{guan2022large} extends graph neural architecture search to handle large-scale graphs containing billions of nodes and edges.

To the best of our knowledge, this work is the first to explore graph neural architecture search for cross-domain recommendation.

\section{The Proposed Method}
In this section, we present our proposed method, a comprehensive and adaptable solution for cross-domain recommendation through behavior importance-aware graph neural architecture search, applicable to both \textbf{dual-domain} and \textbf{multi-domain} settings. The framework is illustrated in Figure~\ref{fig:framework}.

\subsection{Problem Formulation}
\label{sec:formulation}

Before introducing our model, we provide a concise overview of the cross-domain recommendation problem. We consider a scenario with \(N\) users and \(M\) items, where the user-item interaction data is represented as a heterogeneous graph \(\mathcal{G} = (U, I, E)\). Here, \(U = \{u_1, u_2, \ldots, u_N\}\) denotes the set of users, \(I = \{i_1, i_2, \ldots, i_M\}\) the set of items, and \(E \subseteq U \times I\) the set of user-item interactions. Each interaction \((u, i) \in E\) is labeled with a domain \(d \in \mathcal{D}\), where \(\mathcal{D}\) represents the set of domains, and a click label \(y_e\), where 1 indicates a click and 0 indicates no click.

Although recommendation involves multiple stages (e.g., candidate generation, ranking), this paper primarily focuses on the click-through rate (CTR) prediction task, in line with previous works~\cite{ning2023multi, ouyang2020minet}. The goal of cross-domain recommendation is to predict a user's preference for items in the target domain $d_T$, leveraging supplementary information from one or more source domains $d_S \in \mathcal{D}$, where $d_S \neq d_T$. Given interaction data $\mathcal{G}_{d_S}$ from the source domain and $\mathcal{G}_{d_T}$ from the target domain, the task is to learn a function $f: U \times I \times \mathcal{D} \rightarrow \mathbb{R}$ that estimates the likelihood of interaction between user $u \in U$ and item $i \in I$ in the target domain $d_T$.

In the dual-domain setting, where there is only one source domain, this is referred to as dual-domain cross-domain recommendation. In the multi-domain setting, where multiple source domains are involved, the function $f$ must effectively integrate information from multiple source domains $\{d_{S1}, d_{S2}, \ldots, d_{Sk}\}$ to improve prediction accuracy in the target domain $d_T$, where $k$ represents the number of source domains.

It is worth noting that our method adapts to both dual-domain and multi-domain CDR, optimizing data importance and architecture to improve target domain recommendations regardless of the source domain count.

\subsection{Cross-Domain Graph Neural Architecture Search}
\label{sec:nas}

The Cross-Domain Customized Supernetwork module aims to identify the optimal graph neural network architecture for each domain. In this paper, we employ a one-shot graph neural architecture search method that shares weights through a supernet~\cite{liu2018darts}. The optimization objectives are defined as:
\begin{equation}
    \begin{aligned}
        a^*          & = \text{argmax}_{a\in\mathcal{A}}\text{AUC}_{\text{valid}}(a, \mathcal{G}, \mathbf{w}^*),                   \\
        \mathbf{w}^* & = \text{argmin}_{\mathbf{w}}\mathbb{E}_{a\in\mathcal{A}}\mathcal{L}_\text{main}(a,\mathcal{G},\mathbf{w}),
    \end{aligned}
    \label{eq:one-shot}
\end{equation}
where $a$ represents a graph neural network architecture, $\mathcal{A}$ denotes the search space of all possible architectures, and $\mathcal{G} = (U, I, E)$ is a heterogeneous interaction graph constructed solely from training data as described in the problem formulation section. The parameters of the supernetwork are denoted by $\mathbf{w}$, with $\mathbf{w}^*$ representing the optimized parameters. The best-performing architecture within the search space is denoted by $a^*$. AUC, a commonly used evaluation metric for recommendation tasks, will be introduced in the evaluation metric section. The term $\mathcal{L}_\text{main}(a,\mathcal{G},\mathbf{w})$ refers to the loss of architecture $a$ on the training set. Specifically, for the click-through rate prediction (CTR) problem focused on in this paper, the loss can be computed as:
\begin{equation}
    \begin{aligned}
        \mathcal{L}_\text{main}(a,\mathcal{G},\mathbf{w}) & = \mathbb{E}_{e\in E_{\text{train}}}\mathcal{L}(a,\mathcal{G},\mathbf{w}, e), \\
        \mathcal{L}(a,\mathcal{G},\mathbf{w}, e)          & = \text{BCE}(\hat{y}_e, y_e),
    \end{aligned}
\end{equation}
where $\text{BCE}$ denotes the binary cross-entropy loss between the predicted label $\hat{y}_e$ and the true label $y_e$, making it well-suited for click-through rate prediction tasks.

To obtain the optimal architecture $a^*$, we first construct a supernetwork $\mathcal{S}$, following principles from the one-shot NAS literature~\cite{liu2018darts}. The supernetwork is an over-parameterized model that encompasses all possible graph neural network architectures within the search space $\mathcal{A}$, blending multiple operations into a continuous space, represented as:
\begin{align}
    f^{(i)}(\mathbf{x}) = \sum \nolimits_{o\in \mathcal{O}} p_o^i o(\mathbf{x}),
\end{align}
where $\mathbf{x}$ is the input, $f^{(i)}(\mathbf{x})$ is the output, $\mathcal{O}$ is the set of candidate operations, $o \in \mathcal{O}$ is an operation, and $p_o^i$ is the learnable weight of operation $o$ in the $i$-th layer.

The weights $p_o^i$ are optimized via gradient descent. To streamline this process, we maintain a single set of parameters for each operation across the layers. The optimization of the supernetwork is formulated as:
\begin{align}
    \{p_o^i\}^* = \text{argmin}_{\{p_o^i\}} \mathbb{E}_{a\in\mathcal{A}} \mathcal{L}_\text{main}(a, \mathcal{G}, \{p_o^i\}),
\end{align}
where $\mathcal{L}_\text{main}(a, \mathcal{G}, \{p_o^i\})$ is the training loss for architecture $a$ on graph $\mathcal{G}$, parameterized by the weights $\{p_o^i\}$ in the supernetwork.

Unlike most NAS methods that discretize and retrain the selected architecture, our approach retains a continuous architecture, enabling end-to-end training without the need for retraining. This continuous relaxation allows for simultaneous architecture search and training within the supernetwork $\mathcal{S}$, enabling direct optimization of the architecture $a^*$ using gradient-based methods. This approach not only increases flexibility but also streamlines the overall process.

\subsection{Graph-Based Behavior Importance Perceptron}
To optimize the supernetwork, we introduce the Graph-Based Behavior Importance Perceptron, an outer module designed to guide the learning process by evaluating the importance of each interaction in the source domain \(S\). This is achieved by assigning weights to the loss of each sample from the source domain. For illustration, consider two domains, \(S\) and \(T\), where \(T\) is the target domain. The training loss is calculated as:
\begin{equation}
    \mathcal{L}_\text{main} = \sum l(\hat{y}_{Tj}, y_{Tj}) + \sum \gamma_{Si} l(\hat{y}_{Si}, y_{Si}),
    \label{eq:total-loss}
\end{equation}
where \(\mathcal{L}_\text{main}\) represents the total loss for the target domain \(T\). Here, \(y_{Si}\) and \(y_{Tj}\) are the true labels for edges \(e_{Si}\) and \(e_{Tj}\) from domains \(S\) and \(T\), respectively, while \(\hat{y}_{Si}\) and \(\hat{y}_{Tj}\) are the corresponding predicted labels. The term \(\gamma_{Si}\) is the importance weight assigned to interaction \(e_{Si}\) from domain \(S\), and \(l(\hat{y}, y)\) denotes the binary cross-entropy (BCE) loss function.

By setting \(e_{Si} = (u_k, v_{Sl})\), the calculation of \(\gamma_{Si}\) is given by:
\begin{equation}
    \gamma_{Si} = \gamma_S \cdot \sigma(\gamma_{v_{Sl}} \cdot \gamma_{u_k, v_{Sl}}),
    \label{eq:behavior-importance}
\end{equation}
where \(\sigma(\cdot)\) is a normalization function, \(\gamma_S\) is the domain importance weight for domain \(S\), \(\gamma_{v_{Sl}}\) is the global importance weight for item \(v_{Sl}\) across the cross-domain interaction graph \(\mathcal{G}\), and \(\gamma_{u_k, v_{Sl}}\) is the user-specific importance weight for item \(v_{Sl}\) with respect to user \(u_k\).

During training, these weights are multiplied and normalized across the domain, item, and user levels to produce the final importance weight \(\gamma_{Si}\) for each source domain interaction. This weighted loss then contributes to the overall model optimization.

The calculations for the three types of importance weights are as follows:
\begin{itemize}
    \item \(\gamma_S\): The domain importance weight, a scalar parameter continuously updated during training.
    \item \(\gamma_{v_{Sl}}\): The global importance weight for items, computed by processing the item node representation \(\mathbf{h}_{v_{Sl}}\) (obtained through architecture \(a^*\)) via a multi-layer perceptron (MLP).
    \item \(\gamma_{u_k, v_{Sl}}\): The user-specific weight for items, determined by applying a multi-layer GraphSAGE~\cite{hamilton2017inductive} on \(\mathcal{G}_{\text{train}}\). The item representation \(h'_{v_{Sl}}\) and user representation \(h_{u_k}\) are concatenated and further processed by an MLP.
\end{itemize}

In our approach, the Graph-Based Behavior Importance Perceptron acts as the outer-level task-data scheduler, while the backbone recommendation model—comprising the Cross-Domain Customized Supernetwork \(\mathcal{S}\) and the click-through rate predictor—serves as the inner-level model. Together, they form a bi-level optimization framework.

The outer-level perceptron is updated using a development dataset derived from the training set through random reorganization. Using the same data for both levels can hinder the outer model's ability to enhance the inner model, potentially leading to the collapse of the weight \(\gamma_S\) in Equation \ref{eq:total-loss}. To prevent this, we apply stochastic gradient descent (SGD) with batch optimization, using distinct batches for each level. This strategy avoids collapse and enables efficient bi-level optimization without needing additional data from the target domain.

To further optimize the outer-level perceptron, we employ implicit gradients~\cite{navon2021auxiliary}, commonly used in bi-level optimization when direct gradient computation is impractical due to complexity or high computational cost.

The bi-level optimization problem in our model's training phase is formulated as:
\begin{equation}
    \begin{gathered}
        \phi^* = \text{argmin}_\phi \mathcal{L}_{\text{dev}}(\theta^*(\phi)), \\
        \text{s.t.} \quad \theta^*(\phi) = \text{argmin}_\theta \mathcal{L}_\text{main}(\theta; \phi),
    \end{gathered}
\end{equation}
where \(\theta\) are the parameters of the inner-level cross-domain recommendation model, \(\phi\) are the parameters of the outer-level perceptron, and \(\mathcal{L}_\text{main}\) is the training loss weighted by the importance weight \(\gamma_{Si}\) calculated in Eq.~\eqref{eq:total-loss} and Eq.~\eqref{eq:behavior-importance}.

The inner-level optimization seeks the parameters \(\theta^*(\phi)\) that minimize \(\mathcal{L}_\text{main}(\theta; \phi)\), representing the optimal parameters of the cross-domain recommendation model for fixed perceptron parameters. As \(\phi\) changes, \(\theta^*\) also changes. The outer-level optimization adjusts \(\phi\) to ensure that the inner-level model \(\theta^*(\phi)\) achieves optimal performance on the development dataset \(\mathcal{G}_{\text{dev}}\) in the target domain, where \(\mathcal{G}_{\text{dev}}\) is derived from reordering and reusing \(\mathcal{G}_{\text{train}}\).

\textbf{Inner-Level Optimization}. The inner-level model minimizes $\mathcal{L}_\text{main}$ with fixed perceptron parameters $\phi$, using standard methods like SGD or Adam.

\textbf{Outer-Level Optimization}. The outer-level optimization is more complex as $\mathcal{L}_\text{dev}$ depends indirectly on $\phi$ through $\theta$. Thus, the gradient $\nabla_\phi \mathcal{L}_{\text{dev}}(\theta(\phi))$ requires the Chain Rule:
\begin{equation}
    \nabla_\phi \mathcal{L}_\text{dev}(\theta^*(\phi)) = \nabla_\theta \mathcal{L}_\text{dev}(\theta^*(\phi)) \nabla_\phi \theta^*(\phi).
\end{equation}
While $\nabla_\theta \mathcal{L}_\text{dev}(\theta^*(\phi))$ can be computed via automatic differentiation, calculating $\nabla_\phi \theta^*(\phi)$ requires:
\begin{equation}
    \nabla_\theta \mathcal{L}_\text{main}(\theta^*(\phi), \phi) = 0.
    \label{eq:implicit-gradient}
\end{equation}
Taking gradients w.r.t. $\phi$, we get:
\begin{equation}
    \nabla_\phi \theta^*(\phi) = -(\nabla^2_\theta \mathcal{L}_\text{main})^{-1} \nabla_\phi \nabla_\theta \mathcal{L}_\text{main},
\end{equation}
where the Hessian inverse $(\nabla^2_\theta \mathcal{L}_\text{main})^{-1}$ is approximated by the K-truncated Neumann series:
\begin{equation}
    (\nabla^2_\theta \mathcal{L}_\text{main})^{-1} \approx \sum^K_{n=0} (\mathbf{I} - \nabla^2_\theta \mathcal{L}_\text{main})^n.
\end{equation}
Thus, the implicit gradient for the outer-level perceptron is:
\begin{equation}
    \nabla_\phi \mathcal{L}_\text{dev} = -\nabla_\theta \mathcal{L}_\text{dev} \cdot \sum_{n=0}^K (\mathbf{I} - \nabla^2_\theta \mathcal{L}_\text{main})^n \cdot \nabla_\phi \nabla_\theta \mathcal{L}_\text{main}.
    \label{eq:implicit}
\end{equation}

This can be efficiently computed using the Vector-Jacobian Product method~\cite{lorraine2020optimizing}.

Using this iterative approach, the outer-level perceptron and the inner-level recommendation model are fine-tuned alternately. After the inner model converges with the current perceptron parameters, the perceptron is updated. Repeating this process throughout training ensures the convergence of both models, allowing \model to jointly optimize the recommendation architecture and behavior data importance.

\subsection{Click-Through Rate Predictor}
The GNN-based cross-domain supernet computes node representations, denoted as $\mathbf{h}$. For click-through rate prediction, we follow methods like NGCF~\cite{wang2019neural} and compute the input for the predictor as $\mathbf{h}^*_{ij} = (\mathbf{h}_{u_i}^{(0)} || \mathbf{h}_{u_i}^{(1)} || \mathbf{h}_{u_i}^{(2)}) || (\mathbf{h}_{v_j}^{(0)} || \mathbf{h}_{v_j}^{(1)} || \mathbf{h}_{v_j}^{(2)})$, where $\mathbf{h}^{(k)}$ represents node embeddings from the $k$-th GNN layer, and $||$ denotes vector concatenation. This strategy preserves both node and multi-order neighbor information, enhancing the model’s ability to capture intricate user-item interaction patterns.

Using $\mathbf{h}^*_{ij}$ as input, a multi-layer MLP predicts the click-through rate, $\hat{y} = \sigma(\text{MLP}(\mathbf{h}^*_{ij}))$, where $\sigma$ is the sigmoid function, ensuring the prediction remains between $(0,1)$.

\begin{table}[t]
    \centering
    \small
        \begin{tabular}{cccccc}
            \toprule
            Task                   & Domain  & Users                  & Items   & Interactions & Density \\
            \midrule
            \multirow{2}{*}{Bo-Mo} & Books   & \multirow{2}{*}{37,387} & 49,273  & 792,314      & 0.043\% \\
                                   & Movies  &                        & 236,530 & 945,028      & 0.011\% \\
            \midrule
            \multirow{2}{*}{Bo-CD} & Books   & \multirow{2}{*}{16,738} & 150,190 & 418,603      & 0.017\% \\
                                   & CDs     &                        & 61,201  & 380,675      & 0.037\% \\
            \midrule
            \multirow{2}{*}{Bo-El} & Books   & \multirow{2}{*}{28,506} & 203,698 & 735,192      & 0.013\% \\
                                   & Elec    &                        & 52,134  & 364,267      & 0.025\% \\
            \midrule
            \multirow{2}{*}{Bo-To} & Books   & \multirow{2}{*}{7,576}  & 117,771 & 317,503      & 0.036\% \\
                                   & Toys    &                        & 11,567  & 84,564       & 0.096\% \\
            \midrule
            \multirow{2}{*}{CD-Cl} & CDs     & \multirow{2}{*}{1,390}  & 17,707  & 27,128       & 0.110\% \\
                                   & Cloth   &                        & 8,074   & 12,312       & 0.110\% \\
            \midrule
            \multirow{2}{*}{CD-Ki} & CDs     & \multirow{2}{*}{2,809}  & 28,253  & 53,995       & 0.068\% \\
                                   & Kitchen &                        & 14,274  & 37,559       & 0.094\% \\
            \midrule
            \multirow{2}{*}{El-Cl} & Elec    & \multirow{2}{*}{8,235}  & 31,484  & 99,594       & 0.038\% \\
                                   & Cloth   &                        & 18,703  & 66,470       & 0.043\% \\
            \bottomrule
        \end{tabular}
    
    \caption{Statistics for the Amazon dataset.}\label{tab:statistic_amazon}
\end{table}

\begin{table}[t]
    \centering
    \small
        \begin{tabular}{ccccc}
            \toprule
            Domain & Users & Items & Interactions & Density \\
            \midrule
            Channel & 47,330 & 222,336 & 6,848,744 & 0.065\% \\
            Moments & 47,330 & 150,078 & 6,615,576 & 0.093\% \\
            Discover & 47,330 & 189,928 & 6,539,810 & 0.073\% \\
            \bottomrule
        \end{tabular}
    \caption{Statistics for the WeChat advertising dataset.}
    \label{tab:statistic_industry}
\end{table}

\section{Experiments}
In this section, we report empirical evaluations of our method. First, we introduce the experimental setup. Then, we report the results on benchmark datasets and our collected industry dataset. Finally, we conduct ablation studies and discuss hyper-parameters.

\subsection{Experimental Setup}
\subsubsection{Datasets}
In this paper, we use the Amazon Product 5-core dataset~\cite{mcauley2015image, he2016ups} for \textbf{dual-domain} recommendation due to its broad user interactions across diverse product categories, which makes it a standard choice for cross-domain recommendation (CDR) research. Additionally, we construct an industry advertising dataset from WeChat, a widely used social platform, to explore \textbf{multi-domain} recommendation scenarios.

The Amazon dataset consists of 143M product reviews across 24 categories (e.g., books, clothing, movies) collected between 1996 and 2014, with each category representing a domain. Due to limited user overlap across multiple domains, we focus on dual-domain recommendation tasks, using the same domain pairs and splits as in BIAO~\cite{chen2023crossdomain}.
The industry dataset, sourced from WeChat, contains 20M records across three WeChat features (Channel, Moments and Discover) with shared users. This dataset mitigates the Amazon dataset's limitations in multi-domain tasks by ensuring sufficient data overlap. User data is grouped by traits such as age, gender, and location and hashed for privacy.

Table~\ref{tab:statistic_amazon} and Table~\ref{tab:statistic_industry} summarize the dataset statistics, including the number of users, items, interactions, and interaction density across various domains.

\subsubsection{Baseline Models}
In this paper, we select eight state-of-the-art recommendation models, covering both single- and cross-domain approaches, as our baselines. Since our method builds on a graph neural network (GNN) recommendation algorithm, we compare it with leading GNN-based techniques.

For single-domain recommendation, we include traditional matrix factorization methods such as BPR-MF~\cite{rendle2009bpr} and GNN-based models like NGCF~\cite{wang2019neural} and LightGCN~\cite{he2020lightgcn}.

In the cross-domain category, we evaluate two types of models. The first includes CoNet~\cite{hu2018conet} and MiNet~\cite{ouyang2020minet}, which share information across domains using cross-stitching and attention-based interest balancing. CoNet-B and MiNet-B are BIAO-enhanced variants~\cite{chen2023crossdomain}, leveraging auxiliary learning~\cite{chen2023joint, chen2022auxiliary, chen2022module} to improve information transfer between domains.

We also consider state-of-the-art cross-domain GNN models like HeroGRAPH~\cite{cui2020herograph}, DisenCDR~\cite{cao2022disencdr}, and EDDA~\cite{ning2023multi}, which use advanced techniques such as heterogeneous graph modeling, disentanglement, and domain alignment for superior performance across multiple domains.

\subsubsection{Evaluation Metric}
\label{sec:metric}
We evaluate the CTR task using two widely-used metrics: area under the curve (AUC) and log loss. Additionally, we compute the relative improvement (RelaImpr) for both metrics to measure performance gains over the baseline model.

AUC measures the area under the receiver operating characteristic curve, ranging from 0 to 1, where higher values indicate better performance. Log loss, also known as binary cross-entropy loss, evaluates the accuracy of the predicted probabilities, with lower values indicating better performance.

RelaImpr calculates the relative improvement of the target model over a baseline for both AUC and log loss. For AUC, since random predictions yield 0.5, RelaImpr is computed as $\text{RelaImpr}_{\text{AUC}} = \left( \frac{\text{AUC}_{\text{target}} - 0.5}{\text{AUC}_{\text{baseline}} - 0.5} - 1 \right) \times 100\%$. For log loss, it is defined as $\text{RelaImpr}_{\text{LogLoss}} = \left( \frac{\text{Log Loss}_{\text{baseline}}}{\text{Log Loss}_{\text{target}}} - 1 \right) \times 100\%$.

\subsubsection{Implementation}
All methods are implemented in PyTorch and PyTorch Geometric~\cite{fey2019fast}. The supernetwork consists of two layers, with a GNN architecture search space specifically designed for recommendation tasks, including GCN~\cite{kipf2017semisupervised}, GAT~\cite{veličković2018graph}, GraphSAGE~\cite{hamilton2017inductive}, LightGCN~\cite{he2020lightgcn}, and a Linear layer. 

During training, we employ an early stopping strategy with a maximum of 100 epochs, and a 10-epoch warm-up before bi-level optimization, using Adam as the optimizer. Each method is run with five different random seeds, with performance metrics reported as the average of these runs. To ensure the robustness of the model comparisons, we perform significance tests to rule out the possibility of results occurring by chance.

\subsubsection{Computational Complexity Analysis}
Let $|V|$ and $|E|$ denote the number of nodes and edges in the heterogeneous graph, and $d$ the dimensionality of hidden representations. The time complexity of most GNNs is $O(|E|d + |V|d^2)$. With $\mathcal{O}$ as the set of candidate operations, the time complexity of our graph neural architecture search (GNAS) method becomes $O(|\mathcal{O}|(|E|d + |V|d^2))$. In terms of learnable parameters, the complexity is $O(d^2)$ for most GNNs and $O(|\mathcal{O}|d^2)$ for GNAS. Thus, the use of GNAS only introduces a linear increase in complexity, which adds minimal computational overhead, ensuring that our method remains efficient and scalable.

\begin{table*}[htbp]
    \centering
    \small
        \begin{tabular}{cc|c|ccc|ccccc|cc}
            \hline
            \multicolumn{2}{c|}{\multirow{2}{*}{\textbf{Task}}} & \multirow{2}{*}{\textbf{Metric}} & \multicolumn{3}{c|}{\textbf{Single-Domain Methods}} & \multicolumn{5}{c|}{\textbf{Cross-Domain Methods}} & \multicolumn{2}{c}{\textbf{Ours}} \\
            \cline{4-13}
                                      &                          &                          & BPRMF & NGCF & LightGCN & MiNet-B & CoNet-B & HeroGraph & DisenCDR & EDDA & \textbf{BiGNAS} & RelaImpr                                                                                                                                                                     \\
            \hline
            \multicolumn{1}{c|}{\multirow{14}{*}{\rotatebox{90}{\textbf{Amazon}}}} & \multirowcell{2}{\textbf{Bo-Mo}} & AUC & 0.6799 & 0.7514 & 0.7347 & 0.7639 & 0.7721 & 0.7537 & 0.7653 & \underline{0.7731} & \textbf{0.7795}* & +2.34\%                                                                                                                                                                  \\
                               \multicolumn{1}{c|}{}       &                          & LogLoss                  & 0.5405 & 0.4620 & 0.4844 & 0.4697 & 0.4571 & 0.4649 & 0.4664 & \underline{0.4554} & \textbf{0.4520}* & +3.19\%                                                                                                                                                                  \\
            \cline{2-13}
                               \multicolumn{1}{c|}{}       & \multirowcell{2}{\textbf{Bo-CD}} & AUC & 0.6210 & 0.7024 & 0.6981 & 0.7145 & 0.7274 & 0.7172 & 0.7228 & \underline{0.7291} & \textbf{0.7456}* & +7.20\%                                                                                                                                                                  \\
                                \multicolumn{1}{c|}{}      &                          & LogLoss & 0.5410 & 0.4217 & 0.4308 & 0.3924 & 0.4018 & 0.3950 & 0.3959 & \underline{0.3865} & \textbf{0.3846}* & +0.49\%                                                                                                                                                                  \\
            \cline{2-13}
                                \multicolumn{1}{c|}{}      & \multirowcell{2}{\textbf{Bo-El}} & AUC & 0.6252 & 0.6782 & 0.6645 & 0.6678 & 0.6864 & 0.6800 & \underline{0.6913} & 0.6814 & \textbf{0.7002}* & +1.29\%                                                                                                                                                                  \\
                                \multicolumn{1}{c|}{}      &                          & LogLoss & 0.6504 & 0.5391 & 0.4928 & 0.5015 & 0.4731 & 0.4696 & \underline{0.4674} & 0.4933 & \textbf{0.4581}* & +2.03\%                                                                                                                                                                  \\
            \cline{2-13}
                                \multicolumn{1}{c|}{}      & \multirowcell{2}{\textbf{Bo-To}} & AUC & 0.6309 & 0.7030 & 0.6836 & 0.6883 & 0.7048 & 0.6902 & 0.7038 & \underline{0.7065} & \textbf{0.7268}* & +9.83\%                                                                                                                                                                  \\
                                \multicolumn{1}{c|}{}      &                          & LogLoss & 0.5667 & 0.4871 & 0.4338 & 0.4425 & 0.4289 & \textbf{0.4166}* & 0.4520 & 0.4400 & \underline{0.4195} & -0.69\%                                                                                                                                                                  \\
            \cline{2-13}
                                \multicolumn{1}{c|}{}      & \multirowcell{2}{\textbf{CD-Cl}} & AUC & 0.5428 & 0.6353 & 0.6562 & 0.6020 & 0.6176 & 0.6602 & \underline{0.6832} & 0.6761 & \textbf{0.6949}* & +1.71\%                                                                                                                                                                  \\
                                \multicolumn{1}{c|}{}      &                          & LogLoss & 0.6924 & 0.5537 & 0.5242 & 0.5813 & 0.5767 & 0.5358 & 0.4745 & \underline{0.4533} & \textbf{0.4268}* & +6.21\%                                                                                                                                                                  \\
            \cline{2-13}
                                \multicolumn{1}{c|}{}      & \multirowcell{2}{\textbf{CD-Ki}} & AUC & 0.6090 & 0.6390 & 0.6681 & 0.6212 & 0.6558 & 0.6745 & 0.6954 & \underline{0.6974} & \textbf{0.7090}* & +5.88\%                                                                                                                                                                  \\
                                \multicolumn{1}{c|}{}      &                          & LogLoss & 0.5652 & 0.4821 & 0.4254 & 0.4381 & 0.4263 & 0.4268 & 0.4177 & \underline{0.4088} & \textbf{0.3968}* & +3.02\%                                                                                                                                                                  \\
            \cline{2-13}
                                \multicolumn{1}{c|}{}      & \multirowcell{2}{\textbf{El-Cl}} & AUC & 0.5694 & 0.6257 & 0.6193 & 0.6099 & 0.6308 & 0.6264 & 0.6388 & \underline{0.6410} & \textbf{0.6559}* & +10.57\%                                                                                                                                                                 \\
                                \multicolumn{1}{c|}{}      &                          & LogLoss & 0.6079 & 0.5143 & 0.5028 & 0.5783 & 0.5261 & \underline{0.4804} & 0.5399 & 0.5073 & \textbf{0.4738}* & +1.40\%                                                                                                                                                                  \\
            \hline
            \multicolumn{1}{c|}{\multirow{6}{*}{\rotatebox{90}{\textbf{Wechat}}}}
                                      & \multirowcell{2}{\textbf{Channel}} & AUC & 0.6734 & 0.7459 & 0.7010 & 0.7547 & 0.7621 & 0.7447 & 0.7792 & \underline{0.7906} & \textbf{0.8146}* & +3.04\%                                                                                                                                                                  \\
                                \multicolumn{1}{c|}{}      &                          & LogLoss & 0.5479 & 0.4843 & 0.4691 & 0.4807 & 0.4452 & 0.4765 & 0.4470 & \underline{0.4315} & \textbf{0.4207}* & +2.57\%                                                                                                                                                                  \\
            \cline{2-13}
                                \multicolumn{1}{c|}{}      & \multirowcell{2}{\textbf{Moments}} & AUC & 0.6315 & 0.6804 & 0.6817 & 0.7349 & 0.7596 & 0.7225 & \underline{0.7662} & 0.7643 & \textbf{0.7862}* & +2.61\%                                                                                                                                                                  \\
                                \multicolumn{1}{c|}{}      &                          & LogLoss & 0.5931 & 0.5206 & 0.4943 & 0.4800 & 0.4758 & 0.5371 & 0.4889 & \underline{0.4667} & \textbf{0.4461}* & +4.63\%                                                                                                                                                                  \\
            \cline{2-13}
                                \multicolumn{1}{c|}{}      & \multirowcell{2}{\textbf{Discover}} & AUC & 0.6653 & 0.6993 & 0.7324 & 0.7415 & 0.7673 & 0.7328 & 0.7517 & \underline{0.7723} & \textbf{0.8079}* & +4.61\%                                                                                                                                                                  \\
                                \multicolumn{1}{c|}{}      &                          & LogLoss & 0.5043 & 0.4799 & 0.4854 & 0.4810 & 0.4317 & 0.4621 & 0.4353 & \underline{0.4264} & \textbf{0.4053}* & +5.21\%                                                                                                                                                                  \\
            \hline
        \end{tabular}
    
    \caption{Overall experimental results of our model \model and the baselines models. The best results are in \textbf{bold}, while the second best results are \underline{underlined}. We executed each method with 5 random seeds and presented the mean performance. The superscript * indicates the results of a paired t-test at the 0.05 significance level, comparing our approach against the strongest baseline methods. RelaImpr represents the relative improvement over the best-performing baseline.}
    \label{tab:experiment}
\end{table*}

\subsection{Experimental Results}
Table~\ref{tab:experiment} presents the experimental results of our \model and baseline models on both the Amazon and the WeChat advertising datasets. The key observations are as follows:

\model achieves the highest performance across most tasks and datasets on both metrics, with AUC improvements of up to 10.57\% on the Amazon dataset and 4.61\% on the industry dataset compared to the best baseline. Notably, \model performs better in tasks with limited data overlap, such as CD-Cl and EL-Cl, than in tasks with more overlap (e.g., Bo-Mo, Bo-CD). This suggests that while baseline models capture patterns effectively in data-rich tasks, \modelns's flexibility excels in sparser scenarios. 

We select AUC as the main metric and report the test results of the model with the highest AUC score on the validation set. Consequently, some inconsistencies are observed between the improvements in AUC and LogLoss. Compared to the strongest baseline, EDDA, \model consistently demonstrates superior performance. EDDA prioritizes similarity in embeddings across domains but neglects domain-specific variability. In contrast, \model leverages both domain-specific and shared information, resulting in better performance.

Furthermore, CDR methods consistently outperform single-domain models across all ten tasks. Even though single-domain models are trained with the full data from both source and target domains, their inability to transfer information between domains limits their effectiveness, leading to inferior performance compared to CDR methods.

Lastly, GNN-based methods, such as DisenCDR and EDDA, outperform MLP-based methods like MiNet and CoNet. This is likely due to GNNs' ability to capture higher-order dependencies through neighbor aggregation, making them better suited for uncovering latent user preferences. This further justifies our decision to incorporate graph-based learning in \model.

\subsection{Ablation Study}
In this section, we conduct an ablation study to assess the impact of individual components in our framework. The model variants compared are as follows:
\begin{itemize}[]
    \item \textbf{MANUAL}: Uses a fixed, optimal manually designed GNN architecture applied uniformly across all domains, without dynamic customization.

    \item \textbf{MIX}: Removes the Cross-Domain Customized Supernetwork, assigning equal weights to all operations, to assess the effect of architecture customization.

    \item \textbf{DISCRETE}: Replaces continuous relaxation with discrete architectures, following DARTS~\cite{liu2018darts}, to compare continuous versus discrete optimization.

    \item \textbf{NO-SOURCE}: Excludes source domain data, evaluating the role of cross-domain information for target domain recommendations.

    \item \textbf{NO-IMPO}: Removes the Graph-Based Behavior Importance Perceptron, treating all source domain user behaviors equally, to assess the impact of dynamic interaction weighting.
\end{itemize}

\begin{figure}[htbp]
    \centering
    \includegraphics[width=.9\linewidth]{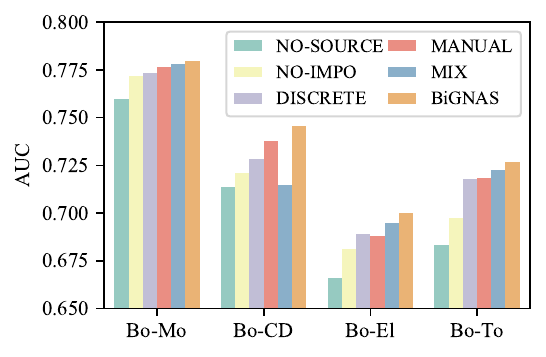}
    \caption{The recommendation performance of different variants of \modelns.}
    \label{fig:ablation}
\end{figure}

We evaluate the model variants on four tasks from the Amazon dataset, with results summarized in Figure~\ref{fig:ablation}. Our complete \model consistently outperforms all alternatives across all tasks, underscoring the importance of each component in cross-domain recommendation. Notably, removing source domain information (NO-SOURCE) results in a significant average performance drop of 5.8\%, while removing the Graph-Based Behavioral Importance Perceptron (NO-IMPO) causes a smaller decline of 4.3\%. These findings highlight the critical role of both modules.

The MIX variant, which incorporates all candidate operations without customizing the GNN architecture, consistently achieves superior performance compared to most other variants, highlighting the effectiveness of integrating multiple message-passing layers to enhance recommendation quality. In contrast, the MANUAL variant, which utilizes the best manually designed architecture, delivers relatively suboptimal results. However, it is noteworthy that MANUAL still outperforms the DISCRETE variant, which relies on discrete architectures generated through traditional NAS methods. This difference in performance is likely attributed to the training approach of the supernet. The DISCRETE variant faces retraining inconsistencies between validation and test sets, while \model resolves this by continuously optimizing the architecture without retraining, ensuring consistent performance.

\section{Conclusion}
In this paper, we propose \modelns, a framework that jointly optimizes GNN architecture and data importance for cross-domain recommendation. Our method automatically customizes the optimal GNN architecture for each domain and dynamically assesses the importance of user behaviors from the source domain. Extensive experiments on benchmark and real-world datasets demonstrate that \model consistently outperforms state-of-the-art baselines in both dual-domain and multi-domain settings, validating its effectiveness for cross-domain recommendation.

\clearpage

\section{Acknowledgments}
This work was supported by the National Key Research and Development Program of China No. 2023YFF1205001, National Natural Science Foundation of China No. 62222209, Beijing National Research Center for Information Science and Technology under Grant No. BNR2023TD03006, BNR2023RC01003, and Beijing Key Lab of Networked Multimedia.

\bibliography{aaai25}

\end{document}